\documentclass[prl,twocolumn,showpacs,amsmath,floatfix]{revtex4}
\usepackage{graphicx}
\usepackage{bm}

\begin{document}

\title{Activation barrier scaling and crossover for noise-induced switching in a micromechanical parametric oscillator}

\author{H. B. Chan}
\email{hochan@phys.ufl.edu}
\author{C. Stambaugh}

\affiliation{Department of Physics, University of Florida, Gainesville, Florida 32611, USA}

%
\begin{abstract}
We explore fluctuation-induced switching in a parametrically-driven micromechanical torsional oscillator. The oscillator possesses one, two or three stable attractors depending on the modulation frequency. Noise induces transitions between the coexisting attractors. Near the bifurcation points, the activation barriers are found to have a power law dependence on frequency detuning with critical exponents that are in agreement with predicted universal scaling relationships. At large detuning, we observe a crossover to a different power law dependence with an exponent that is device specific.
\end{abstract}

\maketitle

Fluctuation-induced escape from a metastable state plays an important role in many physical and biological phenomena. As shown in Kramers' early work for systems in thermal equilibrium \cite{1}, the rate of escape depends exponentially on the ratio of an activation energy to the temperature. Activated transitions have been studied in depth for systems in thermal equilibrium. Recently, there is much interest for activated transitions in systems that are far from equilibrium. A number of systems, such as electrons in Penning traps \cite{2}, Josephson junctions \cite{3}, micro and nanomechanical oscillators \cite{4,5} and atoms in magneto-optical traps \cite{6,7}, develop multistability under sufficiently strong periodic driving, but are monostable otherwise. These systems are not characterized by free energy and the transition rate must be calculated from system dynamics \cite{8,9,10,11}.

In both equilibrium and non-equilibrium systems, as a system parameter $\mu$  approaches a bifurcation value  $\mu_b$, the activation barrier decreases to zero and the number of stable states of the system changes.  In general, the activation barrier is determined by the device parameters and depends on the specifics of the system under study. However, at parameter values close to the bifurcation point, the activation barrier is expected to exhibit system-independent scaling. The activation barrier is proportional to $\left|\mu-\mu_b\right|^{\xi}$  with a critical exponent $\xi$  that is system independent and depends only on the type of bifurcation \cite{8}. For instance, in a Duffing oscillator resonantly driven into bistability, spinodal bifurcations occur at the boundaries of the bistable region. One stable state merges with the unstable state while the other stable state remains far away in phase space. Recent experiments in micromechanical oscillators \cite{5} and rf-driven Josephson junctions \cite{12} have confirmed the theoretical prediction \cite{8,13} that the activation barrier scales with critical exponent 3/2 near spinodal bifurcations in driven systems. On the other hand, a different critical exponent of 2 is expected at a pitchfork bifurcation in systems where all three states merge \cite{14}. Such bifurcation commonly takes place in parametrically driven systems where period doubling occurs. For instance, fluctuation-induced phase slips were observed in parametrically driven electrons in Penning traps \cite{2} between two coexisting attractors and transitions between three attractors were studied in modulated magneto-optical traps \cite{6}. To our knowledge, the activation barriers have not been measured over a wide enough parameter range in these parametrically driven systems to demonstrate the universal scaling at driving frequencies near the two critical points and the crossover to system-specific dependence at large frequency detuning.  

In this Letter, we report measurements of the activation barrier for fluctuation-induced switching in a parametrically-driven micromechanical torsional oscillator, a system that is far from thermal equilibrium. The spring constant of our device is modulated electrostatically near twice the natural frequency. Under sufficiently strong parametric modulation, two pitchfork bifurcation points exist. At the supercritical bifurcation, there emerge two stable oscillation states that differ in phase by  $\pi$.  At the subcritical bifurcation, an additional, stable state with zero oscillation amplitude appears. Noise induces transitions between the coexisting attractors. By measuring the rate of random transitions as a function of noise intensity, we deduce the activation barrier for switching out of each attractor as a function of frequency detuning. Near both bifurcation points, the activation barriers are found to depend on frequency detuning with critical exponent of 2, consistent with the predicted universal scaling in parametrically driven systems \cite{14}. Away from the immediate vicinity of the bifurcation point, universal scaling relationships for the activation barrier no longer hold. We find that in our parametric oscillator, the dependence of the activation barrier on frequency detuning changes from quadratic to $3/2^{\mathrm{th}}$ power. As the driving frequency increases further, the activation barrier becomes independent of the driving frequency for switching between the two oscillation states. In contrast, the barrier for switching out of the zero-amplitude state increases monotonically with frequency detuning. As a result, the occupation of the zero-amplitude state eventually exceeds the two oscillation states. 

\begin{figure}
\includegraphics[angle=0]{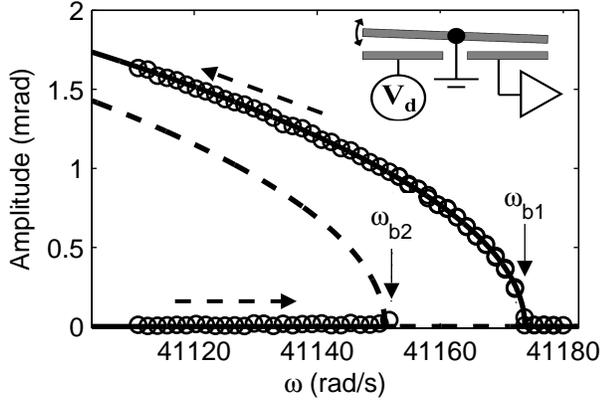} 
\caption{\label{fig:1}Response of the oscillator at $\omega/2$ versus the frequency of parametric driving $\omega$.  The solid and dashed lines represent the stable attractors and the unstable oscillation states respectively.  Inset: cross-sectional schematic of the device (not to scale).}
\end{figure}  
The micromechanical torsional oscillator in our experiment consists of a movable polycrystalline silicon plate ($500 \mu \mathrm{m} \times 500 \mu \mathrm{m}$) suspended by two torsional springs, as shown in the schematic in the inset of Fig. \ref{fig:1}. Two electrodes are located underneath the top plate. A periodic driving voltage  $V_d=V_{dc}+V_{ac}\cos \omega t$ is applied to one of the electrodes, where the driving frequency  $\omega = 2\omega_o + \epsilon$ is close to twice the natural frequency  $\omega_o$(20581.7 $\mathrm{rad\;s}^{-1}$). The top plate is therefore subjected to a periodic electrostatic torque, the angular gradient of which modulates the spring constant. The equation of motion is given by:
 \begin{equation}\label{eq:1}
 \ddot{\theta}+2\gamma \dot{\theta}+\left(\omega_o^2+\frac{k_e}{I}\cos	\omega t \right)\theta+\alpha \theta^2 +\beta \theta^3=0
\end{equation}
where  $k_e=C''(\theta_o)V_{dc}V_{ac}$ is the amplitude of modulation of the spring constant,  $C''(\theta_o)$ is the second derivative of the capacitance between the driving electrode and the top plate with respect to  $\theta$ evaluated at the equilibrium angular position $\theta_o$,  $\gamma$   is the damping constant, $I$ is the moment of inertia,   $\alpha$ and  $\beta$  are the nonlinear coefficients. 
Torsional oscillations of the top plate are detected capacitively by the other electrode. All measurements were performed at 77 K and $< 10^{-7}$ torr. The Q of the oscillator is sufficiently high ($\sim  7500$) so that the response of the oscillator at $\omega \sim 2\omega_o$  is negligible. Details of the oscillator can be found in Ref. [\onlinecite{15}].

When the amplitude of the spring modulation exceeds a threshold value  $k_T=4\omega_o\gamma I$, period doubling occurs \cite{16}. Oscillations are induced at half the modulation frequency in a range close to  $\omega_o$. As shown in Fig. \ref{fig:1}, there are three ranges of frequencies with different number of stable attractors, separated by a supercritical bifurcation point   $\omega_{b1}=2\omega_o+\omega_p$ and a subcritical bifurcation point $\omega_{b2}=2\omega_o-\omega_p$, where  $\omega_p=\sqrt{k_e^2-k_T^2}/2I\omega_o$. In the first region ($\omega>\omega_{b1} \sim 41174\ \mathrm{rad\;s}^{-1}$), no oscillations take place, as the only stable attractor is a zero-amplitude state. At  $\omega_{b1}$, there emerge two stable states of oscillations at frequency  $\omega/2$ that differ in phase by  $\pi$ but are otherwise identical, because of the symmetry with respect to a translation in time by $2 \pi/\omega$. These two stable states are separated in phase space by an unstable state with zero oscillation amplitude (dashed line in Fig. \ref{fig:1}). At frequencies below  $\omega_{b2}(\sim 41150\ \mathrm{rad\;s}^{-1}$), the zero-amplitude state becomes stable, resulting in the coexistence of three stable attractors. These stable states are separated in phase space by two unstable states indicated by the dashed line in Fig. \ref{fig:1}.

The presence of noise allows the oscillator to occasionally overcome the activation barrier and switch between the different attractors. Since the parametrically driven oscillator is far from equilibrium and is not characterized by free energy, the transition rate cannot be determined from the height of the free energy barrier. Theoretical analysis indicates that the transitions remain activated in nature \cite{8}:
\begin{equation}\label{eq:2}
	\Gamma \propto e^{R/I_N}
\end{equation}
where  $\Gamma$ is the transition rate and $I_N$ is the noise intensity. In general, the activation barrier R is determined by the device parameters such as the damping constant, nonlinearity coefficients and the driving frequency. Near the bifurcation points, the system dynamics is characterized by an overdamped soft mode and R decreases to zero according to  $\left|\omega-\omega_b\right|^\xi$, where the critical exponent $\xi$  is universal and depends only on the type of bifurcation. In a parametric oscillator, the supercritical and subcritical bifurcations involve the merging of two stable oscillation states and an unstable zero-amplitude state  (at $\omega_{b1}$) and the merging of two unstable states and a zero-amplitude stable state (at  $\omega_{b2}$) respectively. When three states merge together in such pitchfork bifurcations, the critical exponent is predicted to be 2 \cite{14}. Away from the bifurcation points, the scaling relationship no longer holds and different exponents were obtained depending on the nonlinearity and damping of the system.

  \begin{figure}
\includegraphics[angle=0]{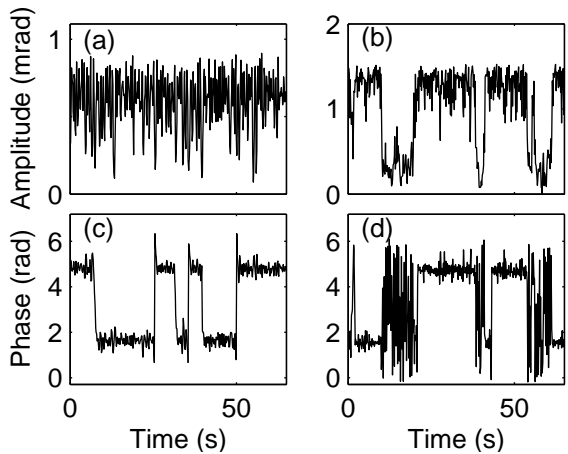} 
\caption{\label{fig:2}Oscillation amplitude (a) and phase (c) for a driving frequency of 41163.0 $\mathrm{rad\;s}^{-1}$. For driving frequencies between $\omega_{b1}$ and  $\omega_{b2}$, transitions occur when the phase slips by  $\pi$. (b) When the driving frequency (41124.8 $\mathrm{rad\;s}^{-1}$) is lower than  $\omega_{b2}$, transitions involve jumps in the amplitude. (d)  In the high amplitude state, the oscillation phase takes on either one of two values that differ by  $\pi$. The phase fluctuates when the oscillator is in the zero-amplitude state.}
\end{figure}
\begin{figure}
\includegraphics[angle=0]{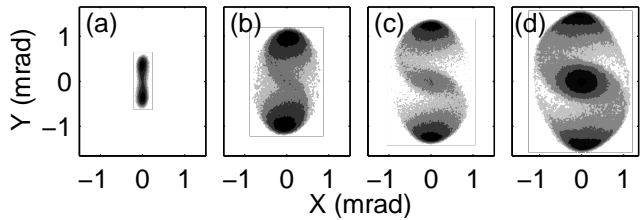} 
\caption{\label{fig:3}The occupation in phase space at four different  $\omega$'s. X and Y denote the two quadratures of oscillation. The grey scale represents the number of times that the oscillator is measured to lie within a certain location in phase space. (a) $\omega$ = 41171.6 $\mathrm{rad\;s}^{-1}$. A pair of oscillation states emerges near  $\omega_{b1}$. (b) $\omega$ = 41153.0 $\mathrm{rad\;s}^{-1}$. As  $\omega$ decreases, the two states move further apart in phase space. (c) $\omega$ = 41139.8 $\mathrm{rad\;s}^{-1}$. When  $\omega<\omega_{b2}$, an addition attractor at the origin appears. (d) $\omega$ = 41124.8 $\mathrm{rad\;s}^{-1}$. With further decrease in  $\omega$, the occupation of the zero-amplitude state increases.}
\end{figure}

  
In order to investigate the transitions between stable states in our parametric oscillator, we inject noise with a bandwidth of $600\ \mathrm{rad\;s}^{-1}$ centered at $\omega_o$. Figures 2a and 2c show respectively the oscillation amplitude and phase at a driving frequency in the range of two coexisting attractors. Transitions can be identified when the phase slips by $\pi$. The two oscillation states have the same amplitude. These two attractors can also be clearly identified in the occupation histogram in Figs. \ref{fig:3}a and \ref{fig:3}b. Figures \ref{fig:2}b and \ref{fig:2}d show switching events at a driving frequency with three attractors, where the zero-amplitude state has also become stable. In contrast to Fig. \ref{fig:2}a, the oscillator switches between two distinct amplitudes. At high amplitude, the phase takes on either one of two values that differ by  $\pi$. When the oscillator is in the zero-amplitude state, there are large fluctuations of the phase as a function of time. The coexistence of three attractors in phase space is also illustrated in Figs. \ref{fig:3}c and \ref{fig:3}d for two other driving frequencies.

\begin{figure}
\includegraphics[angle=0]{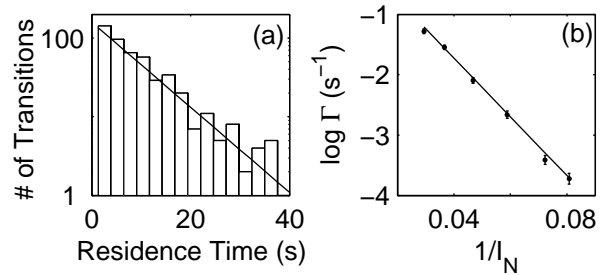} 
\caption{\label{fig:4}Histogram of the residence time in one of the oscillation states before switching occurs, at a driving frequency of 41163.0 $\mathrm{rad\;s}^{-1}$. The solid line is an exponential fit. (b) Logarithm of the transition rate as a function of inverse noise intensity.}
\end{figure}

We identify the residence time in each state before a transition occurs and plot the results as a histogram for one of the oscillation states (Fig. \ref{fig:4}a). The exponential dependence on the residence time indicates that the transitions are random and follow Poisson statistics as expected. From the exponential fit to the histograms, the transition rate out of each state is extracted. The transition rates out of the two oscillation states are measured to be identical to within experimental uncertainty at all noise intensities. Figure \ref{fig:4}b plots the logarithm of the transition rate as a function of the inverse noise intensity. The transition rate varies exponentially with inverse noise intensity, demonstrating that the switching is activated in nature. Using Eq. \ref{eq:2}, we obtain the corresponding activation barriers at a particular driving frequency from the slope in Fig. \ref{fig:4}b. Transitions out of the zero-amplitude state are also found to be activated and follow Poisson statistics in a similar manner.
 
\begin{figure}
\includegraphics[angle=0]{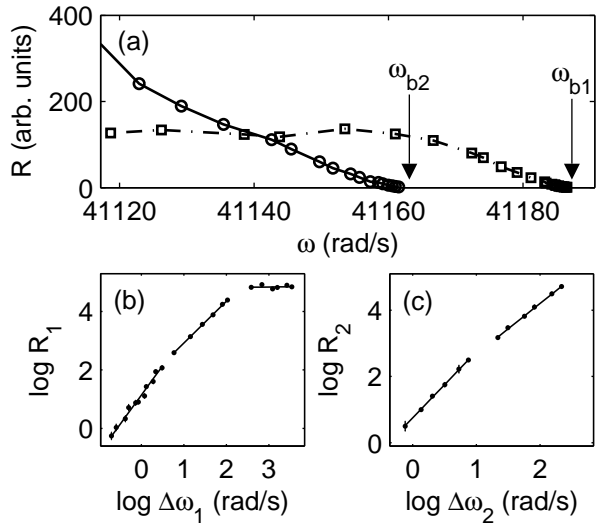} 
\caption{\label{fig:5}Dependence of the activation barriers $\mathrm{R}_1$ (squares) and $\mathrm{R}_2$ (circles) on the frequency of parametric modulation. (b) $\mathrm{R}_1$ vs $\Delta \omega_1$  on a logarithmic scale. The lines are power law fits to different ranges of  $\Delta \omega_1$. (c)  $\mathrm{R}_2$ vs  $\Delta \omega_2$ on a logarithmic scale.}
\end{figure}

The above procedure is repeated to determine the activation barriers at other driving frequencies. Figure 5a shows the driving frequency dependence of the activation barriers $\mathrm{R}_1$ and $\mathrm{R}_2$ for switching out of the oscillation states and the zero-amplitude state respectively. $\mathrm{R}_1$ is calculated using the average transition rates for the two oscillation states. At the high frequency end of Fig. \ref{fig:5}a, only the zero-amplitude state is stable. As the frequency is decreased, two stable oscillation states (separated by an unstable state) emerge at  $\omega_{b1}$. With increasing frequency detuning  $\Delta \omega_1=\omega_{b1}-\omega$, the pair of oscillation states move further apart in phase space (Figs. 3a and 3b) and $\mathrm{R}_1$ increases. At  $\omega_{b2}$, the zero-amplitude state becomes stable. The appearance of the stable zero-amplitude state is accompanied by the creation of two unstable states separating it in phase space from the two stable oscillation states. $\mathrm{R}_2$ initially increases with frequency detuning  $ \Delta \omega_2=\omega_{b2}-\omega$ in a fashion similar to $\mathrm{R}_1$. Close to $\omega_{b2}$, $\mathrm{R}_1$ exceeds $\mathrm{R}_2$ and the occupation of the oscillation states is higher than the zero-amplitude state (Fig. \ref{fig:3}c). As the frequency decreases, $\mathrm{R}_2$ continues to increase monotonically while $\mathrm{R}_1$ remains approximately constant. As a result, $\mathrm{R}_1$ and $\mathrm{R}_2$ cross each other at $\sim 41140\ \mathrm{rad\;s}^{-1}$, beyond which the occupation of the zero-amplitude state becomes higher than the oscillation states. The dependence of the occupation on frequency detuning was also observed in parametrically driven atoms in magneto-optical traps \cite{6}.

In general, the activation barriers $\mathrm{R}_1$ and $\mathrm{R}_2$ depend on various parameters of the device, including the damping constant, nonlinear coefficients, modulation amplitude and frequency. Nonetheless, at frequencies close to the bifurcation points, theoretical analysis indicates that the activation barriers exhibit universal scaling, with $\mathrm{R}_{1,2}=\left|\omega_{b1,2}-\omega \right|^\xi$.  For pitchfork bifurcations in a parametric oscillator that involve merging of three states, $\xi$  is predicted to be 2 \cite{14}. Figures 5b and 5c show the dependence of  $\mathrm{R}_{1,2}$ on frequency detuning $\Delta \omega_{1,2}$  on logarithmic scales.  At small detuning, both activation barriers show power law dependence on detuning. The critical exponents are measured to be 2.0 $\pm$ 0.1 and 2.00 $\pm$ 0.03 for $\mathrm{R}_1$ and $\mathrm{R}_2$ respectively, consistent with theoretical predictions. This quadratic dependence of the activation barrier on detuning near the bifurcation points is system-independent and is expected to occur in other parametrically-driven, nonequilibrium systems such as electrons in Penning traps \cite{2} and atoms in magneto-optical traps \cite{6,7}. Away from the vicinity of the bifurcation point, however, the variation of the activation barrier with frequency detuning is device-specific. Figures 5b and 5c show crossovers from the quadratic dependence to different power law dependence with exponents 1.43 $\pm$ 0.02 and 1.53 $\pm$ 0.02 for $\mathrm{R}_1$ and $\mathrm{R}_2$ respectively. These values are distinct from the exponents obtained in parametrically driven electrons in Penning traps \cite{2} because the nonlinearity and damping are different for the two systems.

Recent theoretical predictions indicate that the symmetry in the occupation of the two oscillation states in a parametrically driven oscillator will be lifted when an additional small drive close to frequency $\omega/2$  is applied \cite{17}. A number of phenomena, including strong dependence of the state populations on the amplitude of the small drive and fluctuation-enhanced frequency mixing, are expected to occur. Further experiments are warranted to test such predictions and reveal other fluctuation phenomena in parametrically driven oscillators.

Parametric pumping is widely used to improve the sensitivity of micromechanical detectors by mechanically amplifying a signal \cite{18} or by reducing the resonance linewidth in viscous environments \cite{20}. The sharp jump in oscillation amplitude at the bifurcation points is utilized for accurate determination of device parameters \cite{4,12}. As the dissipation increases in a resonantly-driven Duffing oscillator, the natural resonance linewidth becomes very broad and the hysteresis region shrinks for comparable oscillation amplitude. Parametrically driven oscillators, on the other hand, maintain the sharp jump in response at the subcritical bifurcation point even for large damping. It is not necessary to increase the oscillation amplitude provided that the stronger parametric pumping compensates the energy loss due to damping \cite{21}. Therefore, parametric oscillators are particular useful for sensing in liquid or gaseous environments. Apart from the relevance to other parametrically driven nonequilibrium systems \cite{2,6,7}, the comprehensive study of the dependence of the transition rate on frequency reported here may prove useful in sensing applications with parametrically driven micromechanical devices.

We thank M. I. Dykman and D. Ryvkine for useful discussions.



\end{document}